\documentstyle[aps,preprint,tighten]{revtex}

\newcommand{\lsim}{\mathrel{\mathop{\kern 0pt \rlap
  {\raise.2ex\hbox{$<$}}}
  \lower.9ex\hbox{\kern-.190em $\sim$}}}
\newcommand{\gsim}{\mathrel{\mathop{\kern 0pt \rlap
  {\raise.2ex\hbox{$>$}}}
  \lower.9ex\hbox{\kern-.190em $\sim$}}}

\begin{document}

\preprint{
\begin{tabular}{r}
DFTT 49/97
\end{tabular}
}

\title{Pinning down neutralino properties from a possible 
modulation signal in WIMP direct search}

\author{
\bf A. Bottino$^{\mbox{a}}$
\footnote{E--mail: bottino@to.infn.it,donato@to.infn.it,
fornengo@to.infn.it,scopel@ezar76.unizar.es},
F. Donato$^{\mbox{a}}$, N. Fornengo$^{\mbox{a}}$, 
S. Scopel$^{\mbox{b}}$\footnote{INFN Post--doctoral Fellow}
\vspace{6mm}
}

\address{
\begin{tabular}{c}
$^{\mbox{a}}$
Dipartimento di Fisica Teorica, Universit\`a di Torino and \\
INFN, Sezione di Torino, Via P. Giuria 1, 10125 Torino, Italy
\\
$^{\mbox{b}}$ Instituto de F\'\i sica Nuclear y Altas Energ\'\i as, \\
Facultad del Ciencias, Universidad de Zaragoza, \\
Plaza de San Francisco s/n, 50009 Zaragoza, Spain
\end{tabular}
}
\date{September 7, 1997}
\maketitle

\begin{abstract}
We analyze the properties of neutralino under the hypothesis that 
some preliminary experimental results of the DAMA/NaI Collaboration  
may be indicative of a yearly modulation effect. We examine which
supersymmetric configurations would be singled out by the DAMA/NaI
data. We also discuss the possibility to investigate these configurations
by means of experimental searches for relic neutralinos other than
direct searches. We finally discuss the possibility to probe these 
configurations by accelerator searches.
\end{abstract}  

\section{Introduction}

In Ref. \cite{nov} it was shown that the DAMA/NaI experiment for 
direct WIMP search, which uses a large-mass, low-background NaI (Tl) 
detector at the Gran Sasso Laboratory \cite{dama}, has currently a 
sensitivity good enough to investigate relic neutralinos in sizeable 
regions of the supersymmetric parameter space.  It was also noticed in 
Ref.\cite{nov} that this is the prerequisite for a significant study of 
a yearly modulation effect, which appears to be the main experimental 
mean available at present for an efficient signal/background 
discrimination in direct relic-particle  detection. Other interesting 
experimental strategies  for disentangling a WIMP signal from the 
background are still at an R\&D stage \cite{sheffield}. 

It is now reported by the DAMA/NaI Collaboration an analysis of a 
collection of data over an exposure of 4549 Kg $\times$ days: 
3363.8 Kg $\times$ days during the winter, 1185.2 Kg $\times$ days during the summer, 
obtained with an experimental set-up
consisting of nine 9.70 Kg NaI(Tl) detectors \cite{dama1}. 
A stability control of the apparatus, based on the $^{210}$Pb peak at 
46.5 KeV, allowed the DAMA/NaI Collaboration to analyse their data in 
terms of a yearly modulation effect. The technique for the extraction 
of a possible signal is a maximum likelihood method applied to a 
binning in the recoil energy of the daily counts per detector. The most 
intriguing result of the investigation carried out in Ref.\cite{dama1} 
is that their set of data appears to be compatible with no-modulation only 
at a 10\% probability level. When interpreted in terms of a modulation 
signal due to a WIMP of mass $m_{\chi}$ and scalar elastic cross 
section (off nucleon) $\sigma^{(\rm nucleon)}_{\rm scalar}$, 
the data of Ref. \cite{dama1} single 
out (at 90\% C.L.) the region of Fig.1, which is delimited by a closed 
contour (hereafter defined as region $R_m$). The open curve in Fig.1 
denotes the 90\% C.L. upper bound of $\sigma^{(\rm nucleon)}_{\rm scalar}$, 
as obtained from the 
total counting rates of Ref. \cite{dama}. Both open and closed contour 
lines were obtained employing for the 
astrophysical parameters the set I of Table 1 (notice that the relevant 
lines of Fig.1 of Ref. \cite{dama1} correspond to the value 
$\rho_l = 0.3$ GeV cm$^{-3}$ for the local dark matter density, instead 
of the value $\rho_l = 0.5$ GeV cm$^{-3}$ adopted here).

As stressed in Ref. \cite{dama1}, the occurrence of region $R_m$ as a domain 
relevant for a possible  modulation effect will require further 
investigation with experimental runs 
with a much higher statistics. Meanwhile, in view of the relevance of 
the issue at stake, we consider of great interest to analyse the 
following questions: a) what would be the features of a 
neutralino (as a definite WIMP candidate) to satisfy the prerequisites 
of region $R_m$; b) would any other experimental 
search for relic neutralinos be able to 
investigate the region $R_m$; c) are neutralino configurations of
region $R_m$ accessible to accelerator searches 
in the near future? The present paper addresses all of these questions. 
Point a) will be analysed in the Minimal Supersymmetric extension of the 
Standard Model (MSSM) \cite{susy}.
As far as point b) is concerned, we consider here the possible signals 
of up--going muons which would be originated by neutralino--neutralino 
annihilation inside the Earth or the Sun \cite{jkg,nupaper,resol}, 
and, as for point c), we will mainly concentrate on the discovery 
potential at LEP and Tevatron.

The plan of this letter is the following. In Section II we present the 
theoretical scheme adopted for our analysis and for our calculation of the 
relevant quantities for direct and indirect detection of relic 
neutralinos.  Section III is devoted to the discussion of the properties
of the neutralino configurations of region $R_m$ which survive the current
bounds from indirect detection. In Section IV we conclude with some final
comments. 

\section{Supersymmetric model}
\label{sec:model}

The WIMP candidate considered in this paper is the neutralino, defined 
as the lowest--mass linear superposition of photino ($\tilde \gamma$),
zino ($\tilde Z$) and the two higgsino states
($\tilde H_1^{\circ}$, $\tilde H_2^{\circ}$)
\begin{equation}
\chi \equiv a_1 \tilde \gamma + a_2 \tilde Z + a_3 \tilde H_1^{\circ}  
+ a_4 \tilde H_2^{\circ}, 
\label{eq:neu}
\end{equation}

In this paper we employ the Minimal Supersymmetric extension of the Standard Model 
MSSM)\cite{susy}. This model is convenient to describe the supersymmetric
phenomenology at the electroweak scale without too strong theoretical assumptions.
Various properties (relic abundances and detection rates) 
of relic neutralinos have been analyzed in the MSSM by 
a number of authors. Some of the most recent papers are given in 
Refs.\cite{nov,jkg,nupaper,resol,ouromega,coann2,noi,bg,bmsheff}.  

The MSSM is based on the same gauge group as the Standard Model
and contains the supersymmetric extension of its particle content. 
In order to give mass both to down-- and up--type quarks and to 
cancel anomalies, two Higgs doublets $H_1$ and $H_2$ are necessary.
As a consequence, the MSSM contains three neutral Higgs fields: two
of them are scalar fields, the other is a pseudoscalar particle.
At the tree level the Higgs sector is specified by two independent parameters:
the mass of one of the physical Higgs fields, which we choose to
be the mass $m_A$ of the neutral pseudoscalar boson, and the ratio of the 
two vacuum expectation values, defined as $\tan\beta\equiv \langle H_2
\rangle/\langle H_1\rangle$.
Once radiative corrections are introduced, the Higgs sector depends
also on the squark masses through loop diagrams.
The other parameters of the model are defined in the superpotential, 
which contains all the Yukawa interactions
and the Higgs--mixing term 
$\mu H_1 H_2$, and  in the soft--breaking
Lagrangian, which contains the trilinear and bilinear  breaking 
parameters and the soft gaugino and scalar mass terms. 

As it stands, the MSSM contains a large number of free parameters. Therefore,
in order to deal with manageable models, it is necessary to introduce some
assumptions which establish relations among the parameters
at the electroweak scale.
The usual conditions, which are also employed here, are the following: 
i) all trilinear parameters are set to zero except those of the third family, 
which are unified to a common value $A$;
ii) all squarks and sleptons soft--mass parameters are taken as 
degenerate: $m_{\tilde l_i} = m_{\tilde q_i} \equiv m_0$, 
iii) the gaugino masses are assumed to unify at $M_{GUT}$, and this implies that
the $U(1)$ and $SU(2)$ gaugino masses are related at the electroweak scale by 
$M_1= (5/3) \tan^2 \theta_W M_2$. 

After these conditions are applied, the supersymmetric parameter space
consists of six independent parameters. We choose them to be: 
$M_2, \mu, \tan\beta, m_A, m_0, A$. We vary these parameters in
the following ranges: $10\;\mbox{GeV} \leq M_2 \leq  500\;\mbox{GeV},\; 
10\;\mbox{GeV} \leq |\mu| \leq  500\;\mbox{GeV},\;
65\;\mbox{GeV} \leq m_A \leq  500\;\mbox{GeV},\; 
100\;\mbox{GeV} \leq m_0 \leq  500\;\mbox{GeV},\;
-3 \leq {\rm A} \leq +3,\;
1.01 \leq \tan \beta \leq 50$. 

In our analysis the supersymmetric parameter space is constrained by
all the experimental limits obtained from accelerators on
supersymmetric and Higgs searches. In particular, the latest data from 
LEP2 on Higgs, neutralino, chargino and sfermion masses are 
used \cite{lep2}. Moreover, the constraints 
due to the $b \rightarrow s + \gamma$ process \cite{alam} are satisfied.
In addition to the experimental limits, we further constrain the
parameter space by requiring that the neutralino is the Lightest 
Supersymmetric Particle (LSP), i.e., regions where the gluino or squarks or 
sleptons are lighter than the neutralino are excluded. This
requirement is necessary for the neutralino to be the WIMP
candidate. Finally, the regions of the parameter space where
the neutralino relic abundance exceeds the cosmological bound, i.e. 
$\Omega_{\chi}h^2 > 1$, are also excluded.

\subsection{Scalar $\chi$--nucleus cross section}

Neutralinos interact with matter both through coherent effects 
\cite{barbieri,griest} and spin--dependent interactions \cite{griest}.
In the present paper we confine ourselves to the coherent effects, since 
these are the only ones which, with the current experimental 
sensitivities, are actually most easily accessible to direct detection.

The pointlike $\chi$--nucleus coherent cross--section is given by
\begin{equation}
\sigma_{C}^0 = \frac {8 G_F^2} {\pi} M_Z^2 \zeta^2 m_{\rm red}^2 A_N^2\, ,
\label{eq:co}
\end{equation}
where $G_F$ is the
Fermi constant, $M_Z$ is the $Z$ boson mass and
$A_N$ and $m_{\rm red}$ are the nucleus mass number and the 
neutralino--nucleus reduced mass, respectively.
The quantity $\zeta$, which depends on the $\chi$--quark couplings 
mediated by Higgs particles and squarks, were originally evaluated
in Refs. \cite{barbieri,griest}. The full expression for $\zeta$ 
together with the values used here for the relevant parameters are 
reported in Ref.\cite{scopel}. 

Due to the structure of Eq.(\ref{eq:co}) an equivalent $\chi$--nucleon 
scalar cross--section may be defined as
\begin{equation}
\sigma^{(\rm nucleon)}_{\rm scalar} =
\left(\frac{1+m_\chi/m_N}{1+m_\chi/m_P}\right)^2
\frac{\sigma^0_{C}}{A_N^{2}}
\label{eq:nucleon}
\end{equation}
where $m_N$ ($m_P$) is the nuclear (proton) mass. The connection between 
Eq.(\ref{eq:nucleon}) and the differential event rate for elastic 
neutralino--nucleus scattering may be found in Ref. \cite{nov}.  

\subsection{Relic abundance and scaling factor $\xi$}
\label{sec:relic}

The relevant quantity, dependent on the supersymmetric parameters, which 
enters in the event rate of direct detection as well as in the indirect 
signals considered in Sect.\ref{sec:direct}, is $\rho_\chi \times 
\sigma^{(\rm nucleon)}_{\rm scalar}$, where $\rho_\chi$ is the 
neutralino local (solar neighbourhood) density. We factorize $\rho_\chi$
as $\rho_\chi = \xi \rho_l$, i.e. in terms of the (total) local dark 
matter density $\rho_l$. Here $\xi$ is calculated in the following way.
For each point of the parameter
space, we take into account the relevant value of the cosmological neutralino
relic density. When $\Omega_\chi h^2$ is larger than a minimal value
$(\Omega h^2)_{\rm min}$, compatible with observational data and with large--scale 
structure calculations, we simply put $\xi=1$.
When $\Omega_\chi h^2$ turns out  to be less than $(\Omega h^2)_{\rm min}$, 
and then the neutralino may only provide a fractional contribution
${\Omega_\chi h^2 / (\Omega h^2)_{\rm min}}$
to $\Omega h^2$, we take $\xi = {\Omega_\chi h^2 / (\Omega h^2)_{\rm min}}$.
The value to be assigned to $(\Omega h^2)_{\rm min}$ is
somewhat arbitrary, in the range 
$0.03 \lsim (\Omega h^2)_{\rm min} \lsim 0.3$.
The values adopted
in our calculations are listed in Table 1.

The neutralino relic abundance is evaluated here as discussed in
Ref.\cite{ouromega}. Possible refinements due to the inclusion of 
coannihilation effects \cite{coann2,coannih} do not appear to be essential here,
since for the neutralino configurations discussed in this paper, the
mass degeneracy of the LSP with the chargino and the next--to--lightest
neutralino is always greater than 15--20\%. Only a few configurations,
where the neutralino is mainly a higgsino, show a mass degeneracy LSP--chargino
of the order of 10\%. However, these configurations do not provide a cosmologically
relevant neutralino, since $\Omega_\chi h^2 \lsim 0.01$. The mass degeneracy
with sfermions is also very marginal and does not modify the main results
of our analysis.

\subsection{Comparison with the experimental data of direct detection}
\label{sec:direct}

We are now in the position to compare the data of Ref.\cite{dama1} with 
our calculation within the MSSM scheme. The comparison is shown in 
Fig.1, where our results are provided in the form of a scatter plot,
obtained by varying the parameters of the supersymmetric model in
the intervals quoted above. To be definite,
only positive values of $\mu$ are displayed in Fig.1, and all the 
discussion which follows refers to this case. We will comment on
configurations belonging to $\mu<0$ in Sect.\ref{sec:properties}.

It is worth reminding that the density of the points 
in the scatter plot has 
no physical meaning, since it depends on the grid employed in the scanning 
of the supersymmetric parameter space. The main significant information 
from the theoretical scatter plot is provided by its contour. Also, 
we wish to point out that the large spread in the values of 
$\xi \sigma^{(\rm nucleon)}_{\rm scalar}$ reflects the ignorance 
in the physical values of the various supersymmetric parameters.

We see that our scatter plot of Fig.1 reaches 
abundantly the curve of the 90\% C.L. bound and also populates the major 
part of the region $R_m$. These features show that: 1) the sensitivity of 
the DAMA/NaI experiment is adequate for a significant exploration of the 
neutralino parameter space; 2) many physical neutralino configurations are 
actually compatible with a modulation effect with values of $m_\chi$ and
$\sigma^{(\rm nucleon)}_{\rm scalar}$ in the region $R_m$.

We will hereafter call $S$ the set of neutralino configurations whose 
representative points fall inside $R_m$. In the following subsection we 
will apply to the set $S$ the current experimental bounds due to 
neutralino--neutralino annihilation inside the Earth and the Sun.

\subsection{$\chi$--$\chi$ annihilation in the Earth and in the Sun}

Signals of up--going muons generated by $\nu_\mu$'s produced in pair 
annihilation of neutralinos in celestial macroscopic bodies
have been analyzed by many authors. Some of 
the most recent papers in this field are given in 
Ref.\cite{jkg,nupaper,resol,bere}.

The up--going muon fluxes $\Phi_\mu$ reported here have been calculated 
with the procedure discussed in Refs. \cite{nupaper,bere}. The values of 
these fluxes from the Earth, $\Phi_\mu^{\rm Earth}$, and from the Sun,
$\Phi_\mu^{\rm Sun}$, for the configurations of set $S$ are displayed in 
Figs.2a--b together with the experimental 90\% C.L. upper bounds of 
Ref.\cite{baksan} (similar bounds are also provided by MACRO 
\cite{macro}.) In Fig.2a the prominent peak around $m_\chi \simeq 60$ GeV 
is due to the enhancement of the elastic $\chi$--nucleus cross section, because 
of the mass matching of $m_\chi$ with the mass of the $^{56}$Fe nucleus, 
which is a dominant element in the Earth composition. 

From Figs.2a--b it turns out that many configurations of set $S$ are 
disfavored by the present bounds on the up--going muon fluxes. 
However, it is remarkable that still a large portion of set $S$ 
survives the indirect search bounds. Fig.3, when compared with Fig.1,
shows how the limits due to $\Phi_\mu^{\rm Earth}$ and $\Phi_\mu^{\rm Sun}$
restrict the neutralino configurations singled out 
by modulation effects in direct search (in Fig.3 all configurations falling
out of region $R_m$ have been dropped).  We will denote by $T$ the set of 
configurations, belonging to set $S$, which 
satisfy the bounds on the muon fluxes $\Phi_\mu$.

\section{Properties of the neutralino allowed states}
\label{sec:properties}

Let us now consider the features of the supersymmetric configurations 
of set $T$. Some of these properties are displayed in Figs.4--7.

First, we discuss the parameters which determine the
neutralino sector, namely $M_2$, $\mu$ and $\tan\beta$. 
In Fig.4 we display the physical region of neutralino configurations
of set $T$. Dots denote configurations with $10 \leq \tan\beta < 50$,
stars denote configurations with $5 \leq \tan\beta < 50$ and,
finally, full circles represent configurations allowed for any value
of $\tan\beta$ ($1.01 \leq \tan\beta < 50$). The dark area is the
region excluded by current LEP data \cite{lep2}. 
We see that, for relatively small values
of $\tan\beta$, the region in the $\mu$--$M_2$ plane which is
compatible with set $T$ is quite restricted. The reason for this
relies on the fact that for small $\tan\beta$ the neutralino--nucleus
interaction is somewhat smaller with respect to higher values of
$\tan\beta$, where the coupling to quarks is enhanced, both for
the Higgs and squark exchanges. Therefore, lower values of
$\tan\beta$ allow the quantity 
$\xi \sigma^{(\rm nucleon)}_{\rm scalar}$ to be compatible with
set $T$ only in the lower edge of the region plotted in Fig.3,
where the neutralino mass is constrained to a relatively small
range. This property can be seen more clearly in Fig.5, 
where the gray area denotes, for different values of $\tan\beta$,
the neutralino mass ranges compatible with set $T$.
The dark area on the left side of the figure is the region
excluded by current LEP data \cite{lep2}. The region on the left of the
vertical solid line around $m_\chi \simeq 50$ GeV is the
region explorable by LEP at $\sqrt{s} = 192$ GeV \cite{yellow}.
We notice that, as far as the neutralino sector
is concerned, LEP will be able to investigate only marginally
the region of parameter space singled out by set $T$.
A better chance to explore the neutralino mass range
compatible with set $T$ is given by the future upgrades
at the Tevatron and by LHC. As an example, the region which
extends up to the vertical dashed line is the mass region
which, under favourable hypothesis, will be possibly
explored by TeV33\cite{tevatron}.

To complement the information given by the previous figure 
on the neutralino
properties for set $T$, we show in Fig.6 the neutralino composition
for different values of $\tan\beta$. The neutralino composition
is parametrized in terms of the fractional 
amount of gaugino fields in the neutralino
mass eigenstate, i.e. $P=a_1^2 + a_2^2$. 
The two solid lines delimit the region of the
neutralino composition explored in our analysis,
because of the adopted grid over the supersymmetric parameters.
We notice that for small values
of $\tan\beta$ the neutralino is constrained around configurations
of maximal mixing. This is again related to the necessity, for
small values of $\tan\beta$, to have a sufficiently strong coupling
to the nucleus, and this happens mainly for maximal mixed states,
both in the case of Higgs-- and of squark--exchanges. For higher values
of $\tan\beta$ and for $P\sim 0.1$, many configurations are
excluded by the constraints on $\Phi_\mu$.

Let us now discuss the properties of the configurations of set $T$ 
with respect to the other supersymmetric parameters not discussed before,
i.e. $m_A$, $m_0$ and $A$. These parameters, together with $\tan\beta$
and $\mu$, determine the masses of the particles exchanged in the
neutralino--nucleus scattering diagrams, which are crucial in
establishing the size of both
direct and indirect detection signals. The most interesting
characteristic of the configurations of set $T$ is shown in
Fig.7, where $\tan\beta$ is plotted against the mass of the
lightest neutral Higgs field $h$. The dark
regions are the regions which are excluded by current
LEP searches \cite{lep2}
(on the left of the plot) or by theoretical arguments
(on the right side).
The gray area corresponds to values of $m_h$ and $\tan\beta$
compatible with the properties of set $T$.
For low values of $\tan\beta$ only light Higgs masses
are possible ($m_h \lsim$ 100 GeV for $\tan\beta \lsim 5$), 
because the small neutralino--nucleus coupling has to be compensated 
by a light exchanged Higgs particle in order to match the required values of
$\xi \sigma^{(\rm nucleon)}_{\rm scalar}$ of set $T$. For larger
values of $\tan\beta$, higher values of $m_h$ are possible.
However, the region of light $m_h$ and $\tan\beta \gsim 20$ is
not compatible with set $T$ because it provides neutrino fluxes
from the Earth and the Sun which exceed the bounds of Fig.2.
In Fig.7 it is also reported the region which will be accessible to LEP
at $\sqrt{s} = 192$ GeV, with a luminosity of 
150 pb$^{-1}$ per experiment \cite{yellow}. A large portion of the region
compatible with the modulation analysis will be covered by the
LEP analysis. In particular, all the region for 
$\tan\beta \lsim 3$ will be analyzed. 
Thus, the discovery potential of LEP2 for configurations
of set $T$ concerns mainly the possible discovery of
a light Higgs boson. However, direct confirmation
of existence of a chargino and/or a neutralino with the required
properties will very likely demand future experiments at
Tevatron and LHC. 

In the previous discussion we have analyzed the characteristics
of the configurations of set $T$ which turn out to be more
interesting, especially in connection with the discovery potential
at the accelerators in the near future. Furthermore, our analysis has 
pointed out other features. We only quote here that the configurations
of set $T$ do not show correlations with respect to the squark
masses in the physical range explored by the present analysis, except
in the case of the stop mass, where masses of the lightest
stop lighter than 400 GeV are preferred. As regards the stop and sbottom
mixing angles, set $T$ is compatible mainly with negative values
around the maximally mixed configurations:
$-0.8 {\rm \;rad} \lsim \theta_{\tilde b} \lsim -0.3 {\rm \;rad}$
for the sbottom mixing angle and 
$\theta_{\tilde t} \lsim -0.8 {\rm \;rad}$ 
for the stop. Only a few configurations are compatible with set $T$
in the case of positive mixing angles, again around the maximal
squark mixing. This property about the squark mixing angles is 
reversed for $\mu<0$: in this case the preferred region is
for positive values of the mixing angles, around the
value of maximal mixing.

Fig.8 shows the neutralino relic abundance $\Omega_\chi h^2$ 
as a function of the neutralino
mass $m_\chi$ for the configurations of set $T$. The horizontal
line denotes the value of $(\Omega h^2)_{\rm min} = 0.03$ which
we are adopting here for the rescaling procedure, as discussed in 
Sect.\ref{sec:relic}. The most interesting feature of Fig.8
is that many configurations provide a sizeable value for the
relic abundance. Some values of the parameters give 
$\Omega_\chi h^2$ close to 1.
It is remarkable that many supersymmetric configurations of set $T$
provide a neutralino with the prerequisites for being
a good dark matter candidate \cite{bere}.

In order to investigate a little further the characteristics
of the configurations of set $T$ which provide sizeable
neutralino relic abundance, we display in the previously
discussed figures the regions where $\Omega_\chi h^2 > 0.1$
(we name as set $V$ the portion of set $T$ which satisfies this 
requirement). In Fig.5, the region contained inside the long dashed line 
represent, for each value of $\tan\beta$, the intervals of $m_\chi$ 
compatible with set $V$. Only values of $\tan\beta \lsim 20$ are
possible, due to the fact that large values of $\tan\beta$ provide
large annihilation cross sections, with the consequence of
reducing the relic abundance. The neutralino mass intervals
compatible with set $V$ appears not to be accessible to LEP
at $\sqrt{s} = 192$ GeV \cite{yellow}, 
but they could be throughly explored by TeV33\cite{tevatron}.

In Fig.6 the information of the neutralino configurations compatible
with set $V$ are reported in the $P$--$\tan\beta$ plane
(regions below the dashed line). For high values
of $\tan\beta$, compositions
which are purer in the gaugino or higgsino sector are favoured, because
they depress the annihilation cross section. On the contrary,
for low values of $\tan\beta$, mixed neutralino configurations are
compatible with cosmologically relevant values of $\Omega_\chi h^2$.
The configurations of set $V$ are also plotted in the
$\tan\beta$--$m_h$ plane in Fig.7. Again, the region
under discussion is the one inside the dashed line. It is
interesting that the cosmologically appealing configurations
could be extensively explored by the LEP runs 
at $\sqrt{s} = 192$ GeV, as far as the Higgs searches are concerned.

Finally, let us remark that
the analysis which has been discussed so far in this
Section refers to positive values of $\mu$. For
$\mu<0$, our analysis has shown that the situation
is, in general, similar to the case of $\mu>0$, 
apart from the already quoted properties of the squark
mixing angles. However, a few important exceptions
are present: i) $\tan\beta$ is constrained to be
greater than approximately 3; ii) the neutralino relic abundance
does not reach values as high as in the $\mu>0$ case and it
is always less than roughly 0.3.

\section{Conclusions}

In the present paper we have analyzed the properties of neutralinos
under the hypothesis that some preliminary experimental results of the 
DAMA/NaI Collaboration \cite{dama1} may be indicative of a yearly 
modulation effect. As stressed by the same Collaboration, this possible 
indication requires confirmation by the collection and analysis of a 
much higher statistically significant set of data, now under way. 
    Motivated by the intriguing possibility that the effect is real and 
due to a supersymmetric relic particle (neutralino), we have examined 
which supersymmetric configurations would be singled out by the 
DAMA/NaI data. 
    We have shown that some of these configurations are excluded by 
current bounds due to up--going muon fluxes from the Sun and the Earth. 
However, most remarkably a large sample of supersymmetric 
configurations relevant for the possible modulation effect are 
compatible with all available experimental data, in ranges of the 
supersymmetric parameters largely accessible to future investigation 
at accelerators. Also, indirect search for relic neutralinos with 
neutrino telescopes of larger exposure will be capable of 
providing most valuable information on this matter. The discovery 
potential of other experimental means, in particular $\bar p/p$ ratio 
in cosmic rays \cite{pbar,gmsheff}, will be analyzed elsewhere \cite{noifuturo}. 
    An important result of the present study is that a number 
of the analyzed supersymmetric configurations would entail a neutralino 
with a sizeable contribution to the cosmological matter density.

    A further word of caution has to be added here. The analysis 
presented in the present paper relies on the use of a specific set for 
the values of the astrophysical parameters (set I of Table 1). This 
set corresponds to the median values for these parameters. Should the 
astrophysical parameters be on the unfavorable side (set II of Table 1), only a 
very limited sample of supersymmetric configurations would be 
compatible with the region $R_m$ singled out by the data of Ref.\cite{dama1}.
 
\vspace{1cm}
{\bf Acknowledgements.}
We thank the DAMA/NaI Collaboration for making their data available to us
before publication. We also express our thanks to R. Bernabei, P. Belli,
A. Incicchitti and D. Prosperi for very fruitful discussions about the
significance of the experimental results examined in the present paper.

\vfill
\eject

\begin{table}
\caption{Values of the astrophysical and cosmological parameters 
relevant to direct and indirect detection rates. 
$V_{\rm r.m.s.}$ denotes the root mean square velocity of the neutralino Maxwellian
velocity distribution in the halo, $V_{\rm esc}$ is the neutralino escape velocity
and $V_{\odot}$ is the velocity of the Sun around the galactic centre; 
$\rho_{\rm l}$ denotes the local dark matter density and $(\Omega h^2)_{\rm min}$ the
minimal value of $\Omega h^2$. The values of set I are the median values of
the various parameters, the values of set II are the extreme values of the
parameters which, within the physical ranges, provide the lowest estimates of 
the detection rates (once the supersymmetric parameters are fixed).
}
\begin{center}
\begin{tabular}{|c|c|c|}   \hline
 &  Set I &  Set II \\ \hline
$V_{\rm r.m.s}(\rm km \cdot s^{-1}$) & 270 & 245 \\ \hline
$V_{\rm esc}(\rm km \cdot s^{-1}$)   & 650 & 450 \\ \hline
$V_{\odot}(\rm km \cdot s^{-1}$) & 232 & 212 \\ \hline
$\rho_{\rm l}(\rm GeV \cdot cm^{-3}$) & 0.5 & 0.2 \\ \hline
$(\Omega h^2)_{\rm min}$            & 0.03 & 0.3 \\ \hline
\end{tabular}
\end{center}
\end{table}

\vfill
\eject

{\bf Figure Captions}
\vspace{1cm}

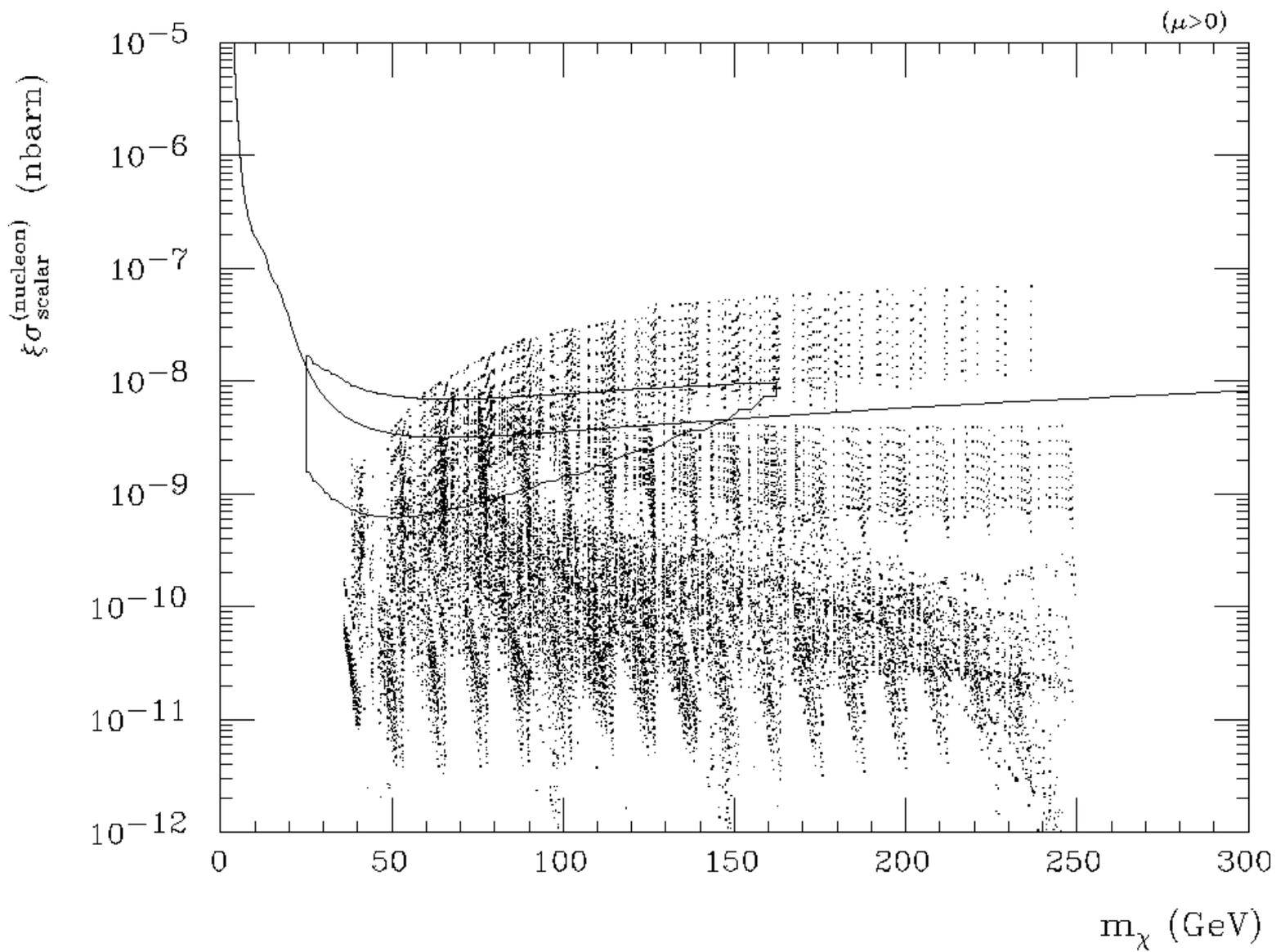
{\bf Figure 1} -- 
The scalar neutralino--nucleon cross section 
$\sigma^{\rm (nucleon)}_{\rm scalar}$, multiplied by the 
scaling factor $\xi$, is plotted versus the neutralino mass $m_\chi$.
The closed contour delimits the region (defined as
$R_m$ in the text)
singled out at 90\% C.L. when the data of ref. \cite{dama1}
are interpreted in terms of a modulation signal.
The open curve denotes the 90\% C.L. upper bound 
obtained from the total
counting rates of Ref.\cite{dama}.
The scatter plot represents the theoretical prediction
for $\xi\sigma^{\rm (nucleon)}_{\rm scalar}$, calculated within the
MSSM scheme. Only configurations with $\mu>$0 are displayed.
The neutralino configurations which fall inside $R_m$
are defined as set $S$. The cosmological and astrophysical 
parameters of set I of Table 1 are used.

{\bf Figure 2a} -- 
Flux of up-going muons $\Phi_{\mu}^{\rm Earth}$
as a function of $m_{\chi}$, calculated
for the neutralino configurations belonging
to set $S$. The solid line represents 
the experimental 90\% C.L. upper bound of 
Ref.\cite{baksan}. Cosmological and 
astrophysical parameters are the same as 
those in Fig.1.

{\bf Figure 2b} -- 
Flux of up-going muons $\Phi_{\mu}^{\rm Sun}$
as a function of $m_{\chi}$, calculated
for the neutralino configurations belonging
to set $S$. The solid line represents 
the experimental 90\% C.L. upper bound of 
Ref.\cite{baksan}.
Cosmological and 
astrophysical parameters are the same as in Fig.1.

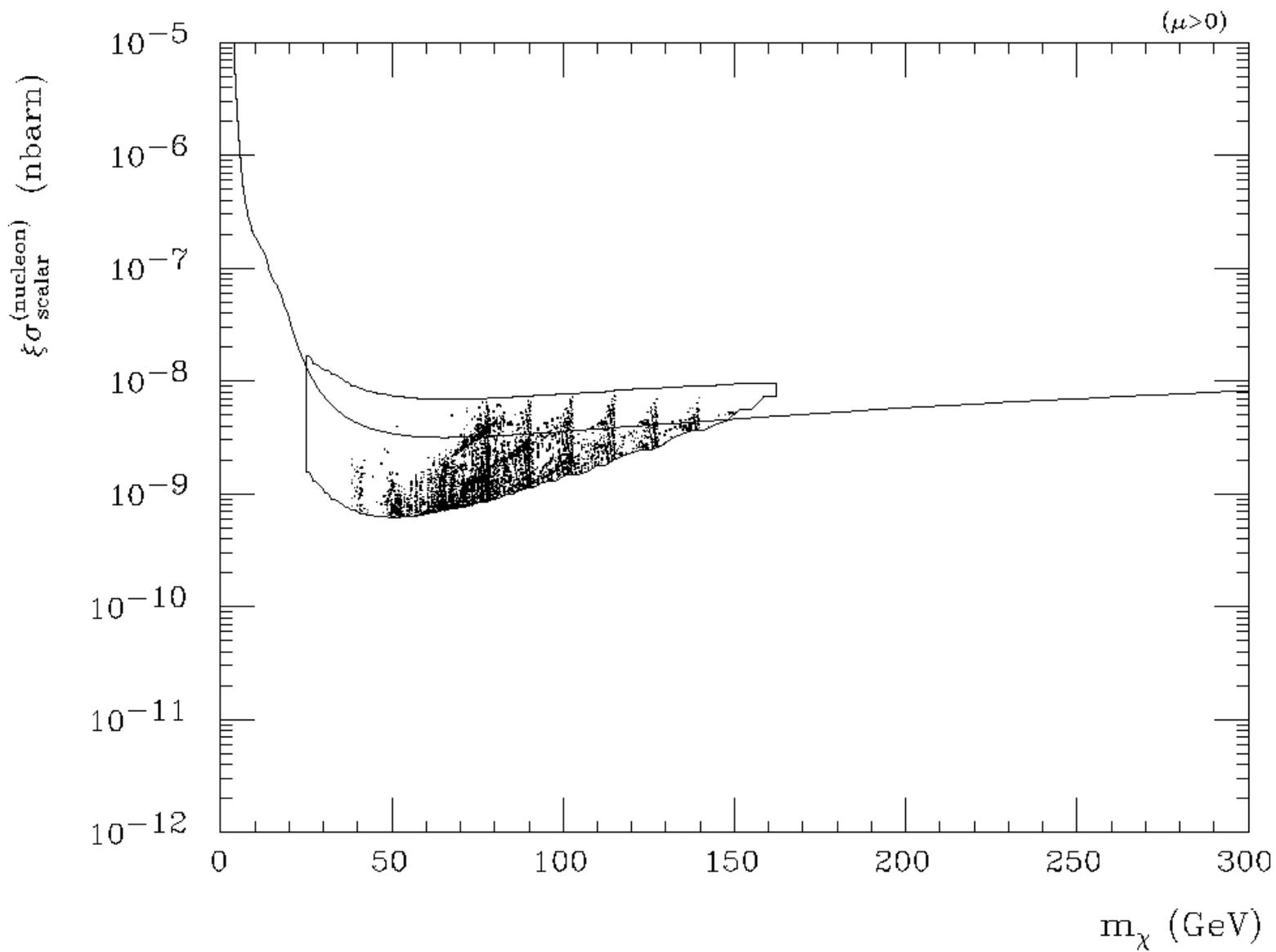
{\bf Figure 3} -- 
Neutralino configurations of set $S$ that survive 
indirect search bounds (referred to as set $T$ in the text).
All configurations falling out of region $R_m$ have been dropped.

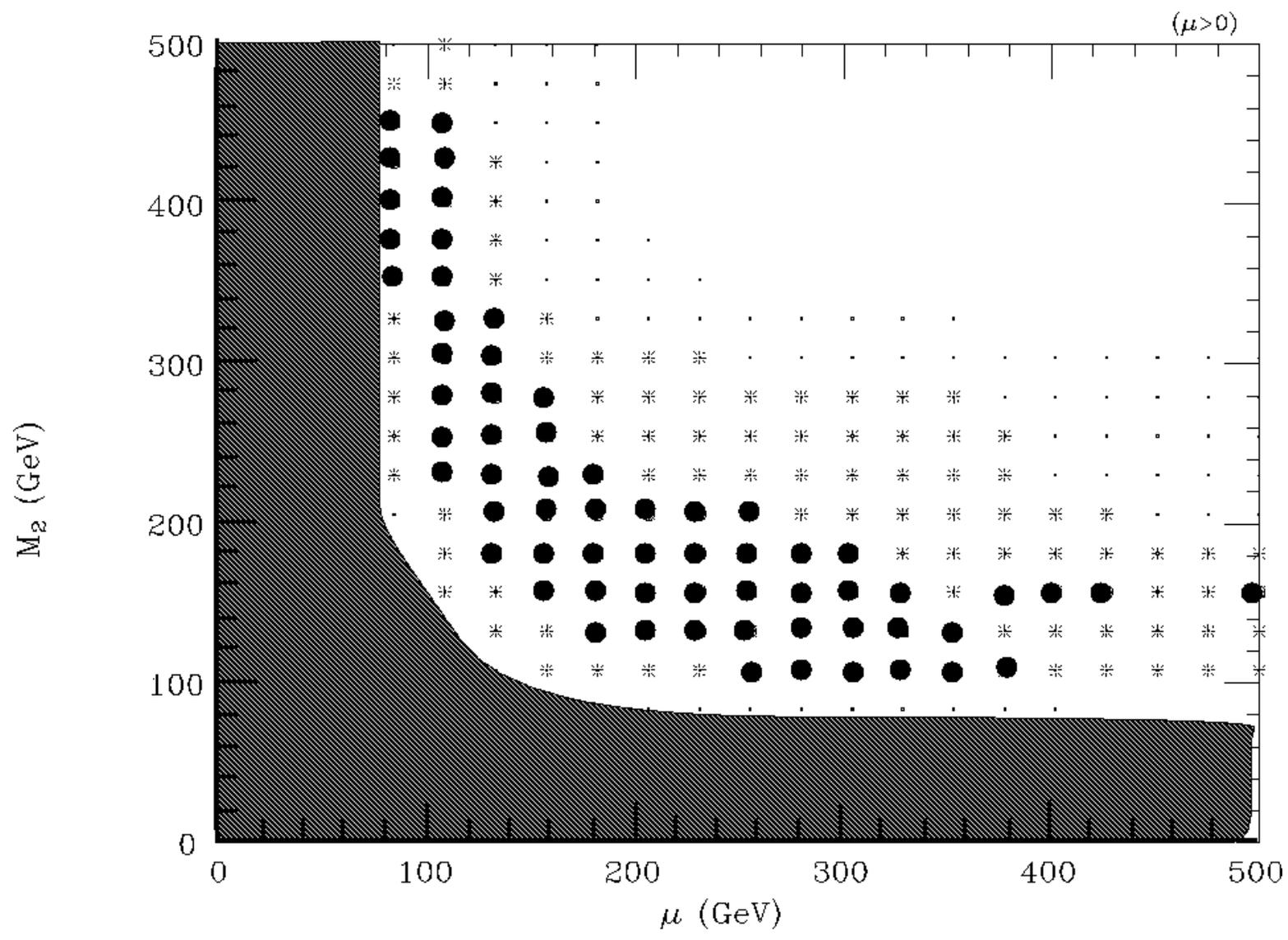
{\bf Figure 4} -- 
Neutralino configurations of set $T$ displayed in the 
$\mu$--$M_2$ plane. The dark area is excluded
by accelerator constraints. Dots denote configurations 
with $10 \leq \tan\beta < 50$;
stars denote configurations with $5 \leq \tan\beta < 50$;
full circles represent configurations allowed for any value
of $\tan\beta$ ($1.01 \leq \tan\beta < 50$).

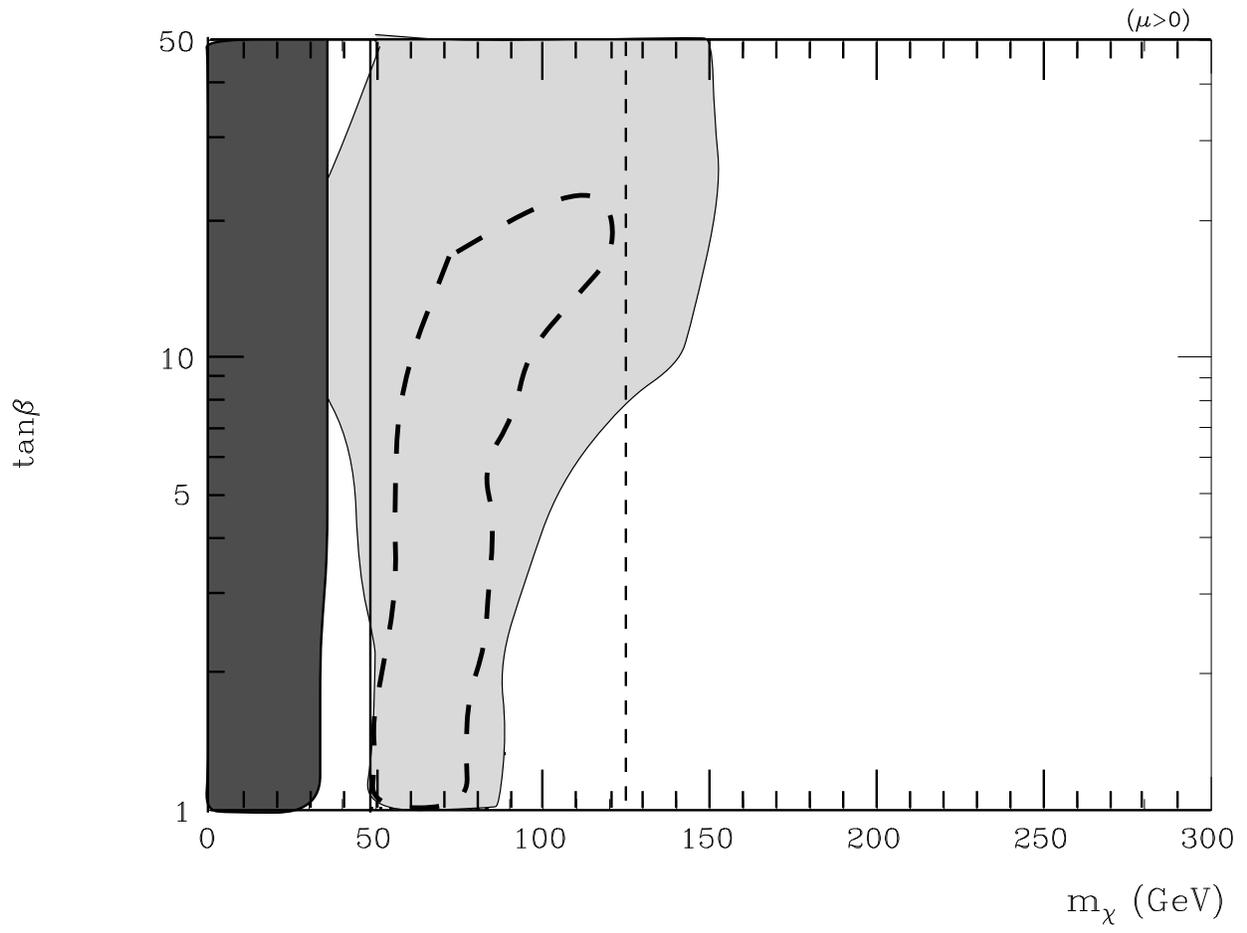
{\bf Figure 5} -- 
The configurations compatible with set $T$ are plotted
in the $m_\chi$--$\tan\beta$ plane, within the gray area. 
The dark region on the left side is 
excluded by current LEP data\cite{lep2}. The region on the left
of the vertical solid line will be accessible to
LEP at $\sqrt{s} = 192$ GeV\cite{yellow}. The region on the left of
the vertical dashed line will be explorable at TeV33\cite{tevatron}.
In the region delimited by the closed dashed line the
neutralino relic abundance $\Omega_\chi h^2$ may exceed {\rm 0.1}. 

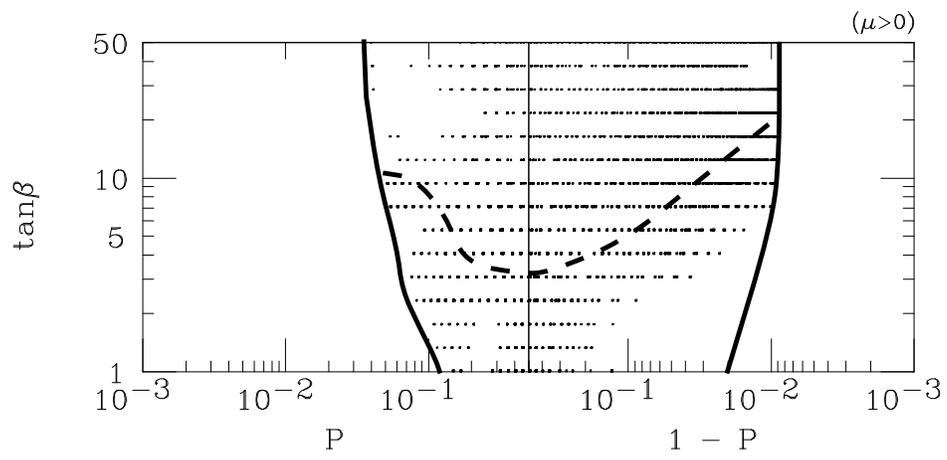
{\bf Figure 6} -- 
The configurations compatible with set $T$ are plotted
in the $P$--$\tan\beta$ plane (dots). On the left
part of the horizontal axis the gaugino fractional amount
$P$ is given, whereas on the right part of the
same axis the complementary variable $1-P$ is reported.
The two solid lines delimit the region of the
neutralino composition explored in our analysis.
Configurations which provide
$\Omega_\chi h^2 > 0.1$ fall below the dashed line.

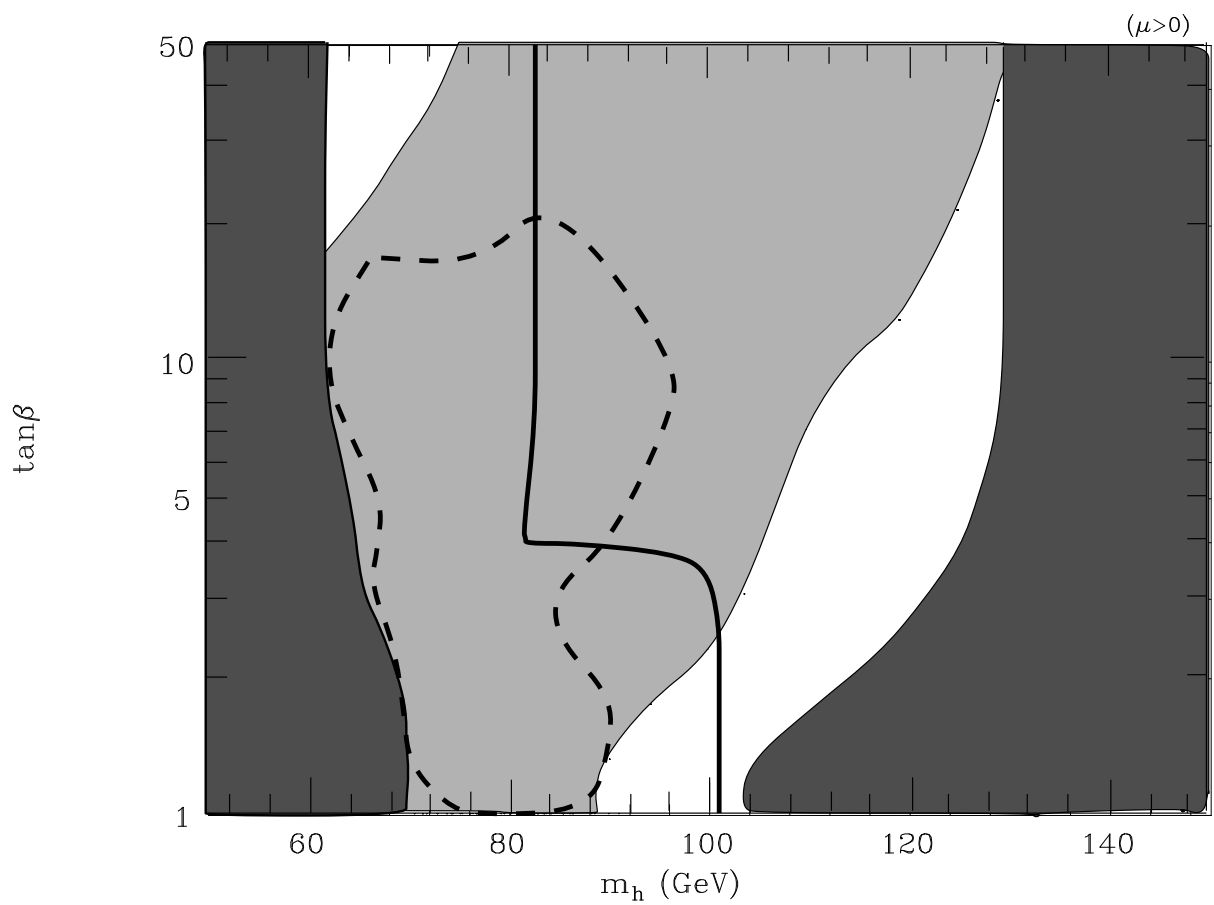
{\bf Figure 7} -- 
The configurations compatible with set $T$ are plotted
in the $m_h$--$\tan\beta$ plane, within the gray area. 
The dark regions are excluded by current
LEP searches \cite{lep2} or by theoretical arguments.
Configurations which provide $\Omega_\chi h^2 > 0.1$
fall within the region delimited by
the closed dashed line.
The region on the left of the solid line 
will be accessible to LEP at $\sqrt{s} = 192$ GeV, 
with a luminosity of 150 pb$^{-1}$ per experiment
\cite{yellow}.

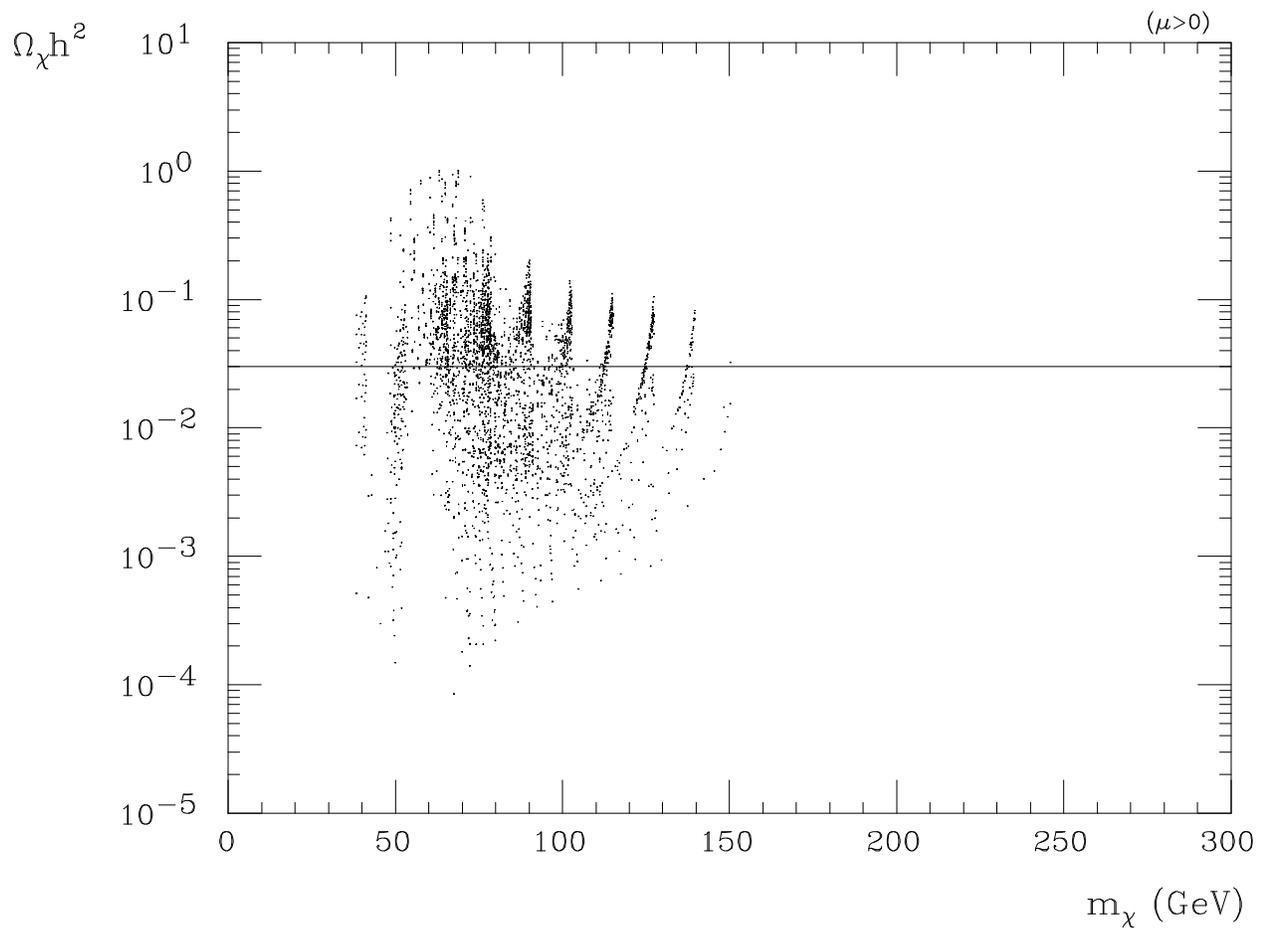
{\bf Figure 8} -- 
Neutralino relic abundance $\Omega_\chi h^2$ as a function
of the neutralino mass $m_\chi$, calculated for the neutralino
configurations of set $T$. The horizontal line denotes
the value of $(\Omega h^2)_{\rm min}$=0.03.

\vfill\eject

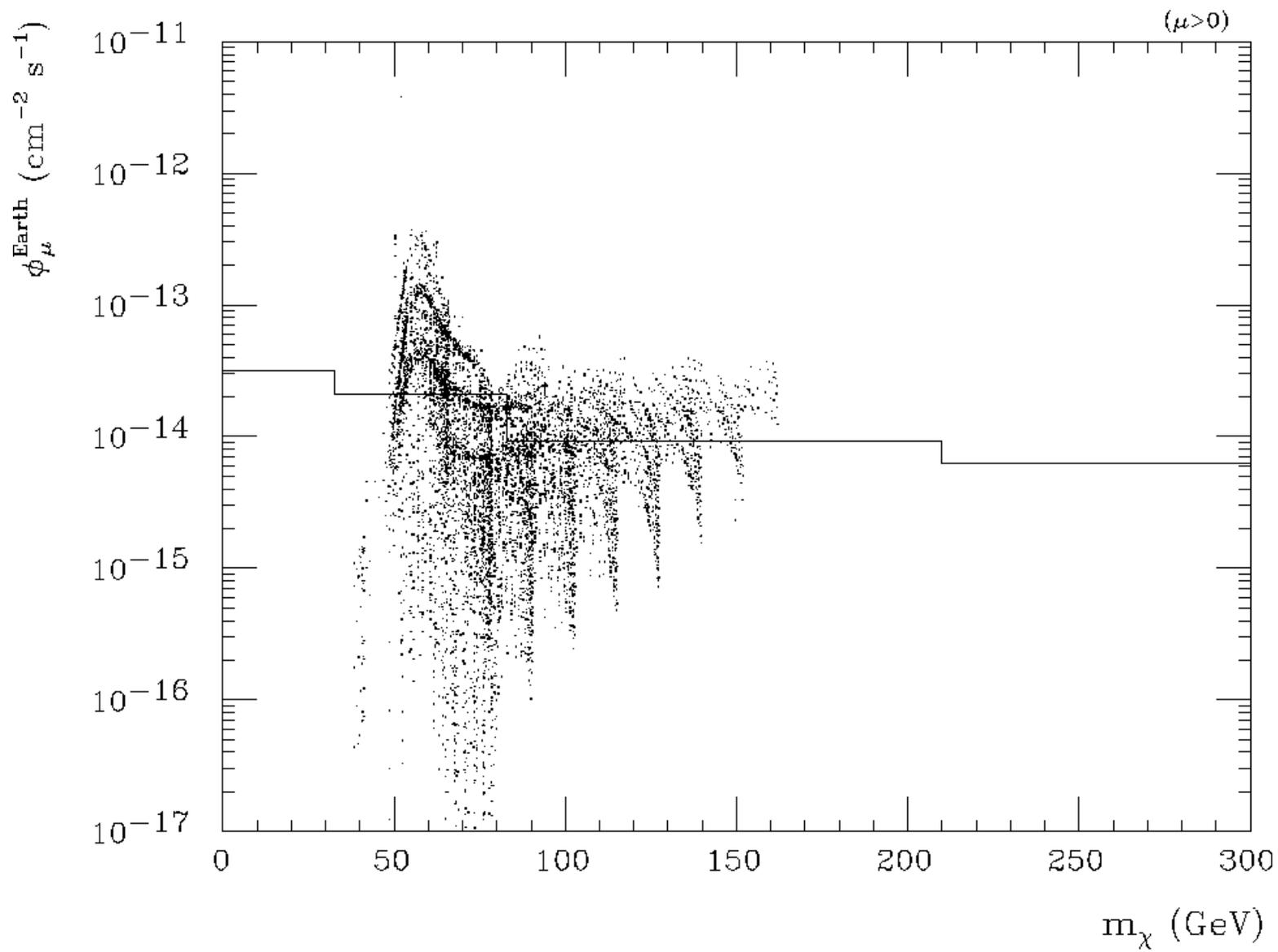
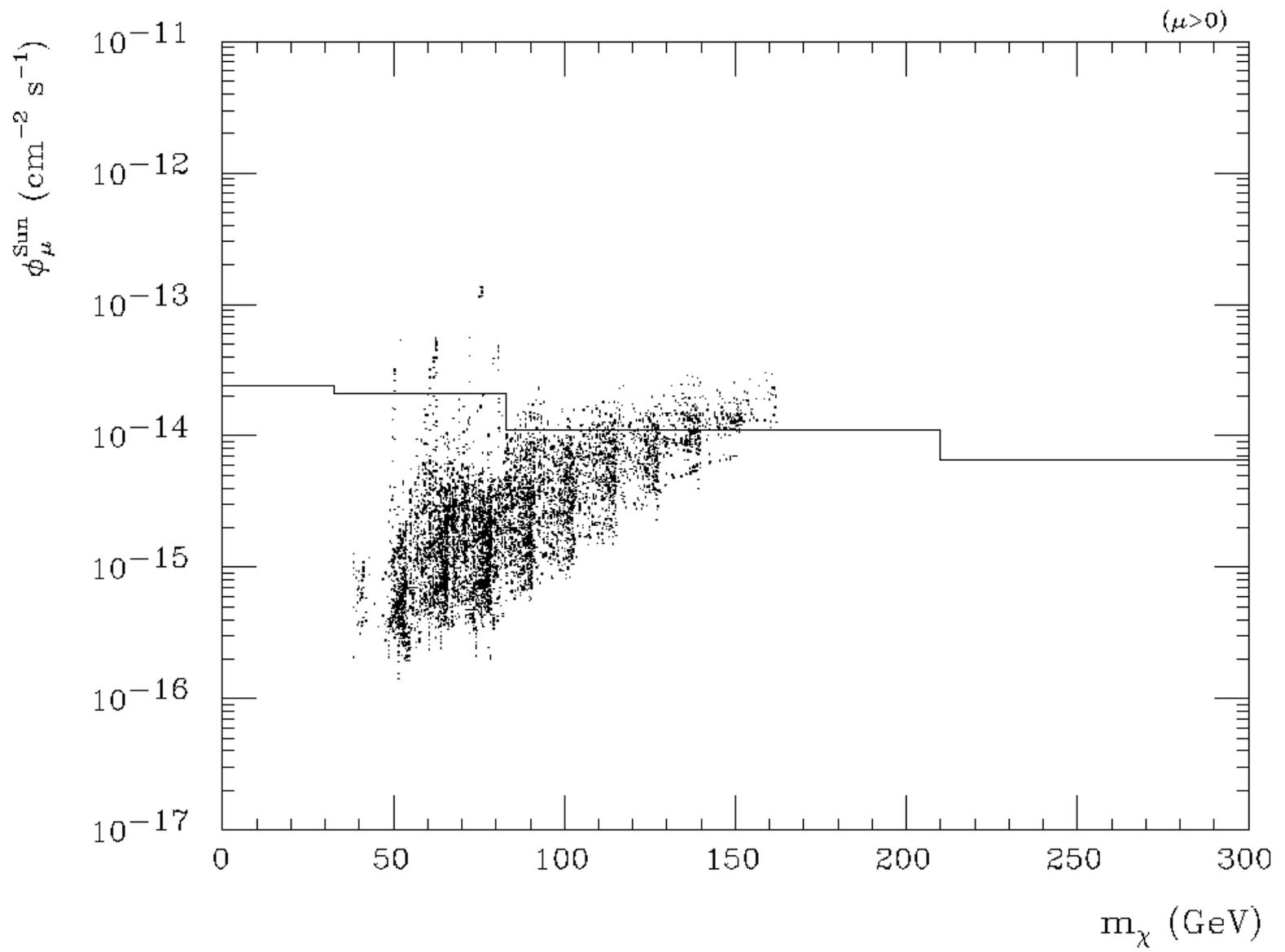
\end{document}